\documentclass[usenatbib, twocolumn]{aastex63}
\usepackage[utf8]{inputenc}
\usepackage{graphicx}
\usepackage{xspace}
\usepackage{comment}
\usepackage{savesym}
\savesymbol{tablenum}
\usepackage{siunitx}
\restoresymbol{SIX}{tablenum}
\usepackage{bm}
\usepackage{url}
\usepackage{footnote}
\usepackage{multirow}
\usepackage{booktabs}
\usepackage{verbatim}
\usepackage{textcomp}
\usepackage{lineno}
\usepackage{float}

\usepackage[xindy, toc, hyperfirst=false, nolist, nostyles, sanitize={name=false,description=false,symbol=false}]{glossaries}
\glsdisablehyper

\newglossaryentry{nlp}{name=NLP, description={Natural Language Processing}, first={Natural Language Processing (NLP)}}

\newglossaryentry{vrad}{name={radial velocity~}, text={radial velocity}, symbol={\ensuremath{v_\textrm{rad}}}, description={radial velocity}, sort=vrad}
\newglossaryentry{vrot}{name={stellar rotation~}, name={stellar rotation}, symbol={\ensuremath{v_\textrm{rot}}}, description={radial velocity}, sort=vrot}

\newcommand{\xray}{X-ray}

\newcommand{\sn}[2]{SN~#1#2\xspace}

\newglossaryentry{angstrom}{name=\AA, description={unit of length $10^{-10}$\,m}, sort=angstrom}
\newglossaryentry{nir}{name=NIR,description={near infrared},first = {near infrared (NIR)}}
\newglossaryentry{psf}{name=PSF,description={point-spread function},first = {point-spread function (PSF)}}
\newglossaryentry{fwhm}{name=FWHM,description={Full Width Half Maximum},first = {FWHM}}
\newglossaryentry{rms}{name=RMS,description={Root Mean Square},first = {RMS}}
\newglossaryentry{signalnoise}{name=S/N,description={signal to noise}}
\newglossaryentry{uv}{name=UV,description={ultra violet},first = {ultra violet (UV)}}
\newglossaryentry{halpha}{name=\ensuremath{\textrm{H}\alpha}, description={First line of the Balmer series at 6563\,\AA}, sort=halpha}
\newglossaryentry{mgb}{name={Mg \textsc{i} b}, description={Triplet at 5167\,\AA, 5173\,\AA and 5184\,\AA}}
\newglossaryentry{sobolevapprox}{name={Sobolev approximation}, description={Lines are approximation with an infinitley thin interaction region \citep[e.g. no broadening][]{1960mes..book.....S}}, first={Sobolev approximation }}
\newglossaryentry{radeq}{name={radiative equilibrium}, description={The net flux of energy between matter and radiation field is zero}}
\newglossaryentry{nebularapprox}{name={nebular approximation}, description={Assumes that the plasma condition are controlled by a central radiation source. The radiation field decreases with the distance to the source by geometrical dilution. See \citet{1978stat.book.....M} for details}}
\newglossaryentry{modnebularapprox}{name={modified nebular approximation}, description={In contrast to \gls{nebularapprox} where only geometrical dilution is taken into account, the modified nebular approximation also takes dilution by other radiative processes into account }, first={modified nebular approximation}, parent=nebularapprox}
\newglossaryentry{thompsonscat}{name={Thomson scattering}, description={Scattering of photons on low energy electrons}}
\newglossaryentry{lte}{name={LTE}, description={Local Thermodynamic Equilibrium}, first={local thermodynamic equilibrium (LTE)}}
\newglossaryentry{lsr}{name={LSR}, description={Local Standard of Rest}, first={\textit{local standard of rest} (LSR)}}
\newglossaryentry{mc}{name={MC}, description={Monte Carlo}, first={\textit{Monte Carlo} (MC)}}
\newglossaryentry{wcs}{name={WCS}, description={world coordinate system}, first={world coordinate system (WCS)}}
\newglossaryentry{cmf}{name=CMF, text=CMF, first=Comoving Frame (CMF henceforth), description={Comoving Frame}}

\newglossaryentry{uvoir}{name=UVOIR, text=UVOIR, first=UV/optical/Near-IR (UVOIR), description={UV/optical/Near-IR}}
\newglossaryentry{ccd}{name=CCD,description={Charged Coupled Device}, first={charged coupled device (CCD)}, firstplural={charged coupled devices (CCDs)}}

\newglossaryentry{ew}{name=Equivalent Width, text={EW}, description={width of a rectangle that has the same area as a spectral line when taken to zero flux}, first={equivalent width (EW)}, firstplural={equivalent widths (EWs)}}
\newglossaryentry{agb}{name=AGB,description={Asymptotic Giant Branch}, first={Asymptotic Giant Branch (AGB)}}
\newglossaryentry{cmb}{name=CMB,description={Cosmic Microwave Background}}
\newglossaryentry{csm}{name=CSM,description={Circumstellar Medium}, first={circumstellar medium (CSM)}}
\newglossaryentry{csi}{name=CSI,description={Circumstellar Interaction}, first={circumstellar interaction (CSI)}}
\newglossaryentry{ism}{name=ISM,description={Interstellar Medium}, first={interstellar medium (ISM)}}
\newglossaryentry{ige}{name=IGE,description={Iron Group Element}, first={iron group element (IGE)}, firstplural={iron group elements (IGEs)}}
\newglossaryentry{epm}{name=EPM,description={Expanding Photosphere Method \citep{1974ApJ...193...27K}}, first={Expanding Photosphere Method (EPM)}}
\newglossaryentry{aic}{name=AIC,description={Accretion Induced Collapse}, first={accretion induced collapse (AIC)}}
\newglossaryentry{ime}{name=IME,description={Intermediate Mass Element}, first={intermediate mass element (IME)}, firstplural={intermediate mass elements (IMEs)}}
\newglossaryentry{h0}{name=\ensuremath{H_0},description={Hubbles constant}}
\newglossaryentry{nse}{name=NSE,description={Nuclear Statistical Equilibrium}, first={nuclear statistical equilibrium (NSE)}}
\newglossaryentry{cdm}{name=CDM,description={Cold Dark Matter}}
\newglossaryentry{grb}{name=GRB,description={Gamma Ray Burst}, first={Gamma Ray Burst (GRB)}, firstplural={Gamma Ray Bursts (GRBs)}}
\newglossaryentry{xps}{name=XPS, description={X-ray point source}, first={X-ray point source (XPS)}, firstplural={X-ray point sources (XPS)}}
\newglossaryentry{donor}{name=donor,description={non-degenerate companion in the \gls{sds}}}
\newglossaryentry{mainsequence}{name=main sequence,description={main sequence star}}
\newglossaryentry{redgiant}{name=red giant,description={red giant star}}
\newglossaryentry{mlcs}{name=MLCS,description={Multicolor Light Curve Shape method \citep[MLCS;][]{1996ApJ...473...88R}}, first={Multicolor Light-Curve Shape method \citep[MLCS;][]{1996ApJ...473...88R}}}
\newglossaryentry{rsoph}{name=RS~Ophiuci ,description={white dwarf accreting from a red giant - assumed progenitor of the \gls{sds}}, sort=rsoph}
\newglossaryentry{usco}{name=U~Scorpii,description={white dwarf accreting from a main sequence star - assumed progenitor of the \gls{sds}}, sort=usco}
\newglossaryentry{rcw86}{name=RCW~86,description={supernova remnant sometimes associated with \sn{185}{}}, sort=rcw86}
\newglossaryentry{casa}{name=Cas~A,description={Cassiopeia A supernova remnant - probably a \gls{snib} event}}
\newglossaryentry{cepheid}{name=Cepheid,description={very luminous variable star with a strong luminosity period relationship}}
\newglossaryentry{urca}{name=Urca, text=\textit{Urca}, description={process predominatly contributing to cooling in stars. The \textit{Urca} process consists of alternating electron-capture and $\beta^{-}$ decay of two nuclei pairs.},sort=urca}
\newglossaryentry{alphacen}{name=Alpha Centauri,description={one of the brightest stars in the night sky and a close binary}}
\newglossaryentry{pcygni}{name={P Cygni}, text={P Cygni},description={a hypergiant luminous blue variable with strong winds. Often referred to as a description for their line profiles showing a emission peak at the rest wavelength of the line and a blue-shifted absorption trough.}}

\newglossaryentry{teff}{name={effective temperature~}, text={effective temperature}, symbol={\ensuremath{T_\textrm{eff}}}, description={Temperature of a blackbody emitting the same total energy}, sort=teff}

\newglossaryentry{logg}{name={surface gravity~}, text={surface gravity}, symbol={\ensuremath{\textrm{log}\,g}}, description={gravity at the surface of a star}, sort=logg}
\newglossaryentry{feh}{name={metallicity~}, text={metallicity}, symbol=\textrm{[Fe/H]},description={iron abundance relative to the sun}, sort=feh}

\newglossaryentry{texp}{name={time since explosion~}, text={time since explosion}, text={time since explosion}, symbol={\ensuremath{t_{\rm exp}}},description={time since explosion (measured in days)}, sort=texp, first={time since explosion (\ensuremath{t_{\rm exp}})}}

\newglossaryentry{lmc}{name=LMC,description={Large Magellanic Cloud}, first={Large Magellanic Cloud (LMC)}, sort=lmc}
\newglossaryentry{smc}{name=SMC,description={Small Magellanic Cloud}, sort=smc}
\newglossaryentry{z}{name=\ensuremath{z},description={redshift}, sort=z}

\newglossaryentry{pca}{name=PCA,description={Princpal Component Analysis}, first={Princpal Component Analysis (PCA)}}

\newglossaryentry{gp}{name=GP,description={Gaussian Process}, first={Gaussian Process (GP)}}

\newglossaryentry{sfit}{name=SFIT, text=\textsc{sfit}, description={spectral fitting program for hot stars \citep{2001A&A...376..497J}}, first={\textsc{sfit} \citep{2001A&A...376..497J}}}
\newglossaryentry{iraf}{name=IRAF, text=\textsc{iraf}, description={Image Reduction and Analysis Facility maintained by NOAO}, first={\textsc{iraf}\protect\footnote{IRAF: the Image Reduction and Analysis Facility is distributed by the National Optical Astronomy Observatory, which is operated by the Association of Universities for Research in Astronomy (AURA) under cooperative agreement with the National Science Foundation (NSF).}}}
\newglossaryentry{pyraf}{name=PyRAF, text=\textsc{PyRAF}, description={Python wrap of \gls{iraf} maintained by STSCI}, first=\textsc{PyRAF} \protect\footnote{PyRAF is a product of the Space Telescope Science Institute, which is operated by AURA for NASA.}}
\newglossaryentry{astropy}{name=ASTROPY, text=\textsc{astropy}, description=\textsc{astropy} framework, first = \textsc{astropy} \citep{2013A&A...558A..33A}}
\newglossaryentry{numpy}{name=NUMPY, text=\textsc{numpy}, description=\textsc{numpy} framework, first = \textsc{numpy} \citep{walt2011numpy}}
\newglossaryentry{scipy}{name=SCIPY, text=\textsc{scipy}, description=\textsc{scipy} framework, first = \textsc{scipy} \citep{Jones:2001fk}}
\newglossaryentry{matplotlib}{name=matplotlib, text=\textsc{matplotlib}, description=\textsc{matplotlib} framework, first = \textsc{matplotlib} \citep{hunter2007matplotlib}}
\newglossaryentry{pandas}{name=pandas, text=\textsc{pandas}, description=\textsc{pandas} framework, first = \textsc{pandas} \citep{mckinney2010data}}
\newglossaryentry{ipython}{name=ipython, text=\textsc{ipython}, description=\textsc{ipython} framework, first = \textsc{ipython} \citep{perez2007ipython}}
\newglossaryentry{jupyter}{name=jupyter, text=\textsc{jupyter}, description=\textsc{jupyter} framework, first = \textsc{jupyter} \citep{kluyver2016jupyter,perez2015project,ragan2014jupyter}}
\newglossaryentry{aplpy}{name=aplpy, text=\textsc{aplpy}, description=\textsc{aplpy} framework, first = \textsc{aplpy} \citep{2012ascl.soft08017R}}
\newglossaryentry{nltk}{name=nltk, text=\textsc{nltk}, description=\textsc{nltk} framework, first = Natural Language ToolKit \citep[\textsc{NLTK};][]{bird2009natural}}
\newglossaryentry{scikit-learn}{name=scikit-learn, text=\textsc{scikit-learn}, description=\textsc{scikit-learn} framework, first = \textsc{scikit-learn} \citep[][]{scikit-learn}}
\newglossaryentry{scikit-image}{name=scikit-image, text=\textsc{scikit-image}, description=\textsc{scikit-image} framework, first = \textsc{scikit-image} \citep[][]{scikit-image}}
\newglossaryentry{moog}{name=MOOG,text={\textsc{moog}}, description={spectral synthesis software \citep{1973ApJ...184..839S}}, first={\textsc{Moog} \citep{1973ApJ...184..839S}}}
\newglossaryentry{atlas9}{name=ATLAS9,description={grid of stellar atmospheres \citep{2004astro.ph..5087C}}, first={ATLAS9 \citep{2004astro.ph..5087C}}}
\newglossaryentry{vald}{name=VALD,description={Vienna Atomic Line Daadtabase \citep{2000BaltA...9..590K}}, first={Vienna Atomic Line Database \citep[VALD;][]{2000BaltA...9..590K}}}
\newglossaryentry{sextractor}{name=SExtractor, text=\textsc{SExtractor}, description={Source Extractor photometry program \citep{1996A&AS..117..393B}}, first={\textsc{SExtractor} \citep{1996A&AS..117..393B}}}
\newglossaryentry{swarp}{name=SWarp, text=\textsc{SWarp}, description={SWarp \citep{2002ASPC..281..228B}}, first={\textsc{SWarp} \citep{2002ASPC..281..228B}}}
\newglossaryentry{astrometry.net}{name=astrometry.net, text=\textsc{astrometry.net}, description={\textsc{astrometry.net} \citep{2010AJ....139.1782L}} first={\textsc{astrometry.net} \citep{2010AJ....139.1782L}}}

\newglossaryentry{astrodrizzle}{name=AstroDrizzle, text=\textsc{AstroDrizzle}, description={AstroDrizzle \citep{2012drzp.book.....G}}, first={\textsc{AstroDrizzle} \citep{2012drzp.book.....G}}}

\newglossaryentry{idl}{name=IDL,text={\textsc{idl}}, description={Interactive Data Language}}
\newglossaryentry{makee}{name=MAKEE,text=\textsc{makee}, description={MAuna Kea Echelle Extraction by Tom Barlow available}}
\newglossaryentry{minuit}{name=MINUIT,text={\textsc{minuit}}, description={collection of numerical optimization tools \citep{James:1975dr}}}
\newglossaryentry{migrad}{name=MIGRAD,text={\textsc{migrad}}, description={numerical gradient optimization tools - part of \gls{minuit}}}
\newglossaryentry{dolphot}{name=DOLPHOT, text=\textsc{dolphot}, description=photometry package for HST, first=\textsc{dolphot} \citep{2000PASP..112.1383D}}
\newglossaryentry{synphot}{name=synphot, text={\textsc{synphot}}, description={synthetic photometry package from STSCI}, first={\textsc{synphot}\protect\footnote{\textsc{synphot} is a product of the Space Telescope Science Institute, which is operated by AURA for NASA.}}}
\newglossaryentry{chianti}{name=CHIANTI, text=CHIANTI, description= CHIANTI Database 7.1, first =CHIANTI 7.1 \citep{1997A&AS..125..149D,2012ApJ...744...99L}}
\newglossaryentry{synpp}{name=SYNPP, text=SYN++, description= SYN++ software, first =SYN++ \citep{2011PASP..123..237T}}
\newglossaryentry{tardis}{name=TARDIS, text=\textsc{tardis}, description= TARDIS MC code, first = {\textsc{tardis} \citep{TARDIS2014}}}

\newglossaryentry{artis}{name=ARTIS, text=\textsc{artis}, description= ARTIS MC code, first = \textsc{artis} \citep{2009MNRAS.398.1809K}}
\newglossaryentry{cmfgen}{name=CMFGEN, text=\textsc{cmfgen}, description=CMFGGEn radiative transfer code, first = \textsc{cmfgen} \citep{1998ApJ...496..407H}}
\newglossaryentry{sedona}{name=SEDONA, text=\textsc{sedona}, description= Sedona MC code, first = \textsc{sedona} \citep{2006ApJ...651..366K}}
\newglossaryentry{phoenix}{name=PHOENIX, text=\textsc{phoenix}, description= PHOENIX radiative transfer code, first = \textsc{phoenix} \citep{1999jcoam.109...41h,1998ApJ...495..370B,1997ApJ...483..390H,1996ApJ...462..386H,1997ApJ...490..803H}}

\newglossaryentry{mesa}{name=MESA, text=\textsc{mesa}, description=Stellar Evolution code, first = \textsc{mesa} \citep{2011ApJS..192....3P,2013ApJS..208....4P,2015ApJS..220...15P,2016ApJS..222....8D,2018ApJS..234...34P,2019ApJS..243...10P}}
\newglossaryentry{stella}{name=STELLA, text=\textsc{stella}, description=Radiative transfer code, first = \textsc{stella} \citep{1998ApJ...496..454B,2004Ap&SS.290...13B,2006A&A...453..229B}}

\newglossaryentry{mlmc}{name=MLMC, text=ML93, description= Mazzali Lucy Monte Carlo, first ={Mazzali \& Lucy (1993, ML93) code}}
\newglossaryentry{starkit}{name=STARKIT, text=\textsc{starkit}, description= TARDIS MC code, first = {\textsc{starkit} \citep{wolfgang_kerzendorf_2015_28016}}}

\newglossaryentry{pyne}{name=PYNE, text=\textsc{pyne}, description= PYNE code, first = {\textsc{pyne} \citep{Scopatz2012a}}}
\newglossaryentry{multinest}{name=MULTINEST, text=\textsc{MultiNest}, description=MultiNest, first={\textsc{MultiNest} \citep{2009MNRAS.398.1601F}}}
\newglossaryentry{wsynphot}{name=WSYNPHOT, text=\textsc{wsynphot}, description=Wsynphot, first={\textsc{wsynphot}\protect\footnote{\protect\url{https://github.com/wkerzendorf/wsynphot}}}}
\newglossaryentry{specutils}{name=SPECUTILS, text=\textsc{specutils}, description=specutils, first={\textsc{specutils} \protect\footnote{\protect\url{https://github.com/astropy/specutils}}}}
\newglossaryentry{ads}{name=ADS ,description=ADS, first={NASA Astrophysics Data System (ADS) \citep{2000A&AS..143...41K}}}

\newglossaryentry{2mass}{name=2MASS,description={Two Micron All Sky Survey \citep{2006AJ....131.1163S}}, first={Two Micron All Sky Survey \citep{2006AJ....131.1163S}}}
\newglossaryentry{wiserep}{name=\textsc{WISeREP}, description={Weizmann Interactive Supernova data REPository \citep{2006AJ....131.1163S}}, first={\textsc{WISeREP} \citep{2012PASP..124..668Y}}}
\newglossaryentry{nomad}{name=NOMAD,first={Naval Observatory Merged Astrometric Dataset \citep[NOMAD; ][]{2005yCat.1297....0Z}}, description={Naval Observatory Merged Astrometric Dataset}}
\newglossaryentry{sdss}{name=SDSS, description={Sloan Digital Sky Survey}}
\newglossaryentry{dss}{name=DSS, description={Digitized Sky Survey}}

\newglossaryentry{eso}{name=ESO, description={European Southern Observatory}, first={European Southern Observatory (ESO)}}
\newglossaryentry{eso.opc}{name=OPC, description={Observing Programmes Comittee}, first={Observing Programmes Comittee (OPC)}}
\newglossaryentry{iau}{name=IAU,description={International Astronomical Union}, first={IAU}}
\newglossaryentry{ctio}{name= CTIO, description={Cerro Tololo Inter-American Observatory}, first={Cerro Tololo Inter-American Observatory (CTIO)}}

\newglossaryentry{nsf}{name=NSF, description={National Science Foundation}, first={National Science Foundation (NSF)}}

\newglossaryentry{wifes}{name=WIFES, text=\textsc{WiFeS}, first={\textsc{WiFeS} \citep{2007Ap&SS.310..255D}},  description={Wide Field Spectrograph - \gls{ifu} mounted on the 2.3\,m telescope at Siding Spring Observatory}}

\newglossaryentry{scp}{name=SCP, description={Supernova Cosmology Project, led by Saul Perlmutter}, first={Supernova Cosmology Project (SCP)}}
\newglossaryentry{hzsns}{name=HZSNS, description={High Z Supernova Search, led by Brian Schmidt}, first={High Z Supernova Search (HZSNS)}}

\newglossaryentry{vlt}{name=VLT,description={Very Large Telescope located on Cerro Paranal (Chile)}, first={Very Large Telescope (VLT)}}
\newglossaryentry{flames}{name=FLAMES,description={Multi-object, intermediate and high resolution spectrograph mounted on the  \gls{vlt}}}

\newglossaryentry{hires}{name=HIRES, description={High Resolution Echelle Spectrometer mounted on the Keck Telescope}, first={High Resolution Echelle Spectrometer \citep[HIRES;][]{1994SPIE.2198..362V}}}

\newglossaryentry{lris}{name=LRIS,description={Low Resolution Imaging Spectrometer mounted on the Keck Telescope}, first={Low-Resolution Imaging Spectrometer \citep[LRIS;][]{Oke95}}}

\newglossaryentry{decam}{name=DECam, description={DECam is a high-performance, wide-field CCD imager mounted at the prime focus of the Blanco 4-m telescope at \gls{ctio}.}, first={Dark Energy Camera \citep[DECam; ][]{2012PhPro..37.1332D,2015AJ....150..150F}}}

\newglossaryentry{essence}{name=ESSENCE,description={The `Equation of State: SupErNovae trace Cosmic Expansion' project \citep[ESSENCE;][]{2002AAS...201.7809G}}, first={`The Equation of State: SupErNovae trace Cosmic Expansion' \citep[ESSENCE;][]{2002AAS...201.7809G}}}
\newglossaryentry{ifu}{name=IFU,description={Optical instrument combining spectrographic and imaging capabilities, used to obtain spatially resolved spectra}, first={Integral Field Unit (IFU)}, firstplural={Integral Field Units (IFUs)}}

\newglossaryentry{besancon}{name=Besan\c{c}on Model, description={Model of stellar population synthesis of the Galaxy, including kinematics.}}

\newglossaryentry{int}{name=INT,description={Isaac Newton 2.5\,m Telescope}, first={Isaac Newton 2.5\,m Telescope (INT)}}

\newglossaryentry{chandra}{name=Chandra,description={Chandra \xray\ Observatory (space-based)}}
\newglossaryentry{hst}{name=HST,description={Hubble Space Telescope}}
\newglossaryentry{hst.wfpc2}{name=WFPC2,description={Wide-Field Planetary Camera 2 mounted on the \gls{hst}}, first={Wide-Field Planetary Camera 2 (WFPC2)}}
\newglossaryentry{hst.acs}{name=ACS,description={Advanced Camera for Surveys mounted on the \gls{hst}}, first={Advanced Camera for Surveys (ACS)}}
\newglossaryentry{hst.wfc3}{name=WFC3,description={Wide-Field Camera 3 mounted on the \gls{hst}}, first={Wide-Field Camera 3 (WFC3)}}
\newglossaryentry{hst.cte}{name=CTE, description={charge transfer efficiency (CTE)}, first={charge transfer efficiency \citep[CTE; see ][for a description]{2009acs..rept....1C}}}

\newglossaryentry{snls}{name=SNLS,description={Supernova Legacy Survey \citep{2003AAS...203.8209P}}, first={Supernova Legacy Survey \citep[SNLS;][]{2003AAS...203.8209P}}}
\newglossaryentry{dass}{name=DASS, description={Digitized Astronomy Supernova Survey \citep{1975PASP...87..565C}}, first={Digitized Astronomy Supernova Survey \citep[DASS;][]{1975PASP...87..565C}}}
\newglossaryentry{bait}{name=BAIT, description={Berkley Automatic Imaging Telescope \citep{1993PASP..105.1164R}}, first={Berkley Automatic Imaging Telescope \citep[BAIT;][]{1993PASP..105.1164R}}}
\newglossaryentry{kait}{name=KAIT, description={Katzman Automatic Imaging Telescope \citep{2001ASPC..246..121F}}, first={Katzman Automatic Imaging Telescope \citep[KAIT;][]{2001ASPC..246..121F}}}
\newglossaryentry{loss}{name=LOSS, description={Lick Observatory Supernova Search  \citep{2000AIPC..522..103L}}, first={Lick Observatory Supernova Search \citep[LOSS;][]{2000AIPC..522..103L}}}
\newglossaryentry{ctss}{name=CTSS,description={Cal\'{a}n/Tololo Supernova Survey \citep{1993AJ....106.2392H}}, first={Cal\'{a}n/Tololo supernova survey \citep[CTSS;][]{1993AJ....106.2392H}}}

\newglossaryentry{ptf}{name=PTF, description={Palomar Transient Factory \citep{2009PASP..121.1334R}}, first={Palomar Transient Factory \citep[PTF;][]{2009PASP..121.1334R}}}
\newglossaryentry{batse}{name=BATSE, description={Burst and Transient Source Experiment mounted on the Compton Gamma Ray Observatory}, first={Burst and Transient Source Experiment (BATSE)}}
\newglossaryentry{bepposax}{name=BeppoSAX, description={\xray\ satellite named in honor of Giuseppe "Beppo" Occhialini}}
\newglossaryentry{rosat}{name=ROSAT, description={short for R\"{o}ntgensatellit}, first={ROSAT}}
\newglossaryentry{hete2}{name=HETE2, description={High Energy Transient Explorer}, first={High Energy Transient Explorer (HETE)}}
\newglossaryentry{ska}{name=SKA, description={Square Kilometre Array}, first={Square Kilometre Array (SKA)}}
\newglossaryentry{swift}{name=Swift, description={Swift Gamma-Ray Burst Mission}}

\newglossaryentry{gnirs}{name=GNIRS, description={Gemini Near InfraRed Spectrograph mounted on the Gemini North Telescope}}
\newglossaryentry{gmosn}{name=GMOS, description={Gemini Multi Object Spectrograph mounted on the
 Gemini North Telescope}, first={GMOS \citep[Gemini Multi Object Spectrograph;][]{2004PASP..116..425H}}}

\newglossaryentry{vla}{name=VLA, description={Very Large Array radio telescope located in North America}, first={Very Large Array (VLA)}}
\newglossaryentry{evla}{name=EVLA, description={Extended Very Large Array radio telescope located in North America}, first={Extended Very Large Array (EVLA)}}
\newglossaryentry{skymapper}{name=SkyMapper, description={SkyMapper telescope \citep{2007PASA...24....1K}}, first={SkyMapper \citep{2007PASA...24....1K}}}
\newglossaryentry{panstarrs}{name=PanSTARRS, description={Panoramic Survey Telescope \& Rapid Response System \citep{2004SPIE.5489...11K}}, first={Panoramic Survey Telescope \& Rapid Response System \citep[PanSTARRS;][]{2004SPIE.5489...11K}}}
\newglossaryentry{ps1dr1}{name=PS1~DR1, description={Panoramic Survey Telescope \& Rapid Response System \citep{2004SPIE.5489...11K} }, first={Panoramic Survey Telescope \& Rapid Response System \citep[PanSTARRS;][]{2004SPIE.5489...11K} DR1}}

\newglossaryentry{lsst}{name=LSST, description={Large Synoptic Survey Telescope}, first={Large Synoptic Survey Telescope \citep[LSST;][]{2006AAS...209.8604P}}}
\newglossaryentry{ppmxl}{name=PPMXL, description={PPMXL Catalog of Positions and Proper Motions on the ICRS \citep{2010AJ....139.2440R}}}
\newglossaryentry{gaia}{name=GAIA, description={Global Astrometric Interferometer for Astrophysics \citep{2001A&A...369..339P}}, first={Global Astrometric Interferometer for Astrophysics \citep[GAIA;][]{2001A&A...369..339P}}}
\newglossaryentry{ligo}{name=LIGO, description={Laser Interferometer Gravitational Wave Observatory}, first={Laser Interferometer Gravitational Wave Observatory \citep[LIGO;][]{1992Sci...256..325A}}}
\newglossaryentry{aligo}{name=Advanced LIGO, description={Advanced LIGO}, sort=ligo2}
\newglossaryentry{lisa}{name=LISA, description={Laser Interferometer Space Antenna \citep{1994ESAJ...18..219J}}, first={Laser Interferometer Space Antenna \citep[LISA;][]{1994ESAJ...18..219J}}}

\newglossaryentry{irc}{name=IRC, text={IRC}, description={infrared catastrophe}, first={infrared catastrophe \citep[IRC;][]{1980PhDT.........1A}}}

\newglossaryentry{sn}{name=Supernova, text={SN}, plural={SNe}, description={exploding star}, nonumberlist=true, first={supernova (SN)}, firstplural={supernovae (SNe)}}
\newglossaryentry{snia}{name=Type~Ia (SN~Ia), text={SN~Ia}, description={Thermonuclear explosion of a white dwarf - spectra show no hydrogen but a strong silicon line},first={Type~Ia supernova (SN~Ia)}, firstplural={Type Ia supernovae (SNe~Ia)}, plural={SNe~Ia}, parent=sn, nonumberlist=true}
\newcommand{\sneia}{\glspl*{snia}\xspace}

\newglossaryentry{branchnormal}{name={branch-normal}, text=\textit{Branch-normal}, description={Large homogeneous class of Type Ia Supernovae, defined in \citet{1993AJ....106.2383B}}, first={\textit{Branch-normal} SNe Ia \citep{1993AJ....106.2383B}}, parent=snia}
\newglossaryentry{91t}{name={91T-like}, description={Luminous class of Type Ia supernovae similar to \sn{1991}{T} \citep{1992AJ....103.1632P}} , first={91T-like}, parent=snia}
\newglossaryentry{91bg}{name={91bg-like}, description={Faint class of Type Ia supernovae similar to \sn{1991}{bg} \citep{1992AJ....104.1543F}}, first={91bg-like}, parent=snia}
\newglossaryentry{02cx}{name={02cx-like}, description={Peculiar class of Type Ia supernovae similar to \sn{2002}{cx} \citep{2003PASP..115..453L}}, first={02cx-like \sneia\ \citep{2003PASP..115..453L}}, parent=snia}

\newglossaryentry{snibc}{name=Type~Ib/c, text={SN~Ib/c}, description={Collapse of the core of a massive star -  spectrum shows no hydrogen and no silicon line},first={Type~Ib/c supernova (SN~Ib/c)}, firstplural={Type~Ib/c supernovae (SNe~Ib/c)}, plural={SNe~Ib/c}, parent=sn}

\newglossaryentry{snib}{name=Type~Ib, text={SN~Ib}, description={Spectrum shows no hydrogen and no silicon, but helium line},first={Type Ib supernova (SN~Ib)}, firstplural={Type~Ib supernovae (SNe~Ib)}, plural={SNe~Ib}, parent=snibc}

\newglossaryentry{snic}{name=Type~Ic, text={SN~Ic}, description={Spectrum shows no hydrogen, no silicon and no helium line},first={Type~Ic supernova (SN~Ic)}, firstplural={Type~Ic supernovae (SNe~Ic)}, plural={SNe~Ic}, parent=snibc}

\newglossaryentry{snii}{name=Type~II, text={SN~II}, description={Collapse of the core of a massive star - spectrum shows strong hydrogen line},first={Type~II supernova (SN~II)}, firstplural={Type~II supernovae (SNe~II)}, plural={SNe~II}, parent=sn}

\newglossaryentry{sniib}{name=Type~IIb, text={SN~IIb}, description={Spectrum shows hydrogen and helium lines},first={Type~IIb supernova (SN~IIb)}, firstplural={Type~IIb supernovae (SNe~IIb)}, plural={SNe~IIb}, parent=snii}

\newglossaryentry{sniip}{name=Type~II~Plateau (Type IIP), text={SN~IIP}, description={Lightcurve shows plateau},first={Type~IIP supernova (SN~IIP)}, firstplural={Type~II Plateau supernovae \citep[SNe~IIP;][]{1979A&A....72..287B}}, plural={SNe~IIP}, parent=snii}

\newglossaryentry{sniil}{name=SN~II~Linear, text={SN~IIL}, description={Lightcurve shows no plateau, but linear decline},first={Type~IIL supernova (SN~IIL)}, firstplural={Type~II~Linear supernovae \citep[SNe~IIL;][]{1990MNRAS.244..269S}}, plural={SNe~IIL}, parent=snii}

\newglossaryentry{sniin}{name=Type II narrow-lined (Type IIn), description={Spectrum shows narrow lines},first={Type~II~narrow-lined supernova (SN IIn)}, firstplural={Type~IIn supernovae (SNe~IIn)}, plural={SNe~IIn}, parent=snii}

\newglossaryentry{snr}{name=Remnant (SNR), text=SNR, description={Remnant left visible post-explosion}, first={supernova remnant (SNR)}, firstplural={supernova remnants (SNRs)}, parent=sn}

\newglossaryentry{dtd}{name=DTD,description={delay time distribution - expected supernova rate over time after a brief outburst of starformation},first={delay time distribution (DTD)}, firstplural={delay time distributions (DTDs)}, plural=DTDs}

\newglossaryentry{hvg}{name=HVG,description={high velocity gradient - Type Ia supernovae with a fast evolution of photospheric velocity},first={high velocity group (HVG)}, firstplural={high velocity groups (HVGs)}, plural=HVGs, parent=snia}

\newglossaryentry{lvg}{name=LVG,description={low velocity gradient - Type Ia supernovae with a slow evolution of photospheric velocity},first={low velocity group (LVG)}, firstplural={low velocity groups (LVGs)}, plural=LVGs, parent=snia}

\newglossaryentry{wd}{name=white dwarf (WD), text=WD, description={White Dwarf - extremely dense stellar remnant}, first={white dwarf (WD)}}
\newglossaryentry{onemgwd}{name= Oxygen/Neon (ONe), text={ONe-WD},description={Oxygen/Neon White Dwarf}, first={oxygen/neon White Dwarf (ONe-WD)}, parent=wd}
\newglossaryentry{cowd}{name=carbon/oxygen (CO), text={CO-WD}, description={carbon/oxygen white dwarf}, first={carbon/oxygen white dwarf (CO-WD)}, firstplural = {carbon/oxygen white dwarfs (CO-WDs)}, parent=wd}

\newglossaryentry{sds}{name=SD-Scenario,description={single-degenerate scenario (single white dwarf accreting from non-degenerate companion)}, first={single-degenerate scenario (SD-scenario)}}

\newglossaryentry{dds}{name=DD-Scenario, description={double degenerate scenario (merging of two white dwarfs)}, first={double-degenerate scenario (DD-scenario)}}

\newglossaryentry{sss}{name=SSS, text={supersoft \xray\ source}, description={supersoft \xray\ source - believed to be emitted by nuclear fusion on a white dwarf's surface}}

\newglossaryentry{amcvn}{name=AM CVn, description={AM Canum Venaticorum star \citep[white dwarf accreting hydrogen poor matter from a companion star; see ][]{2005ASPC..330...27N}}}

\newglossaryentry{rlof}{name=RLOF, description={Roche Lobe Overflow (see \citet{1971ARA&A...9..183P} for a more detailed description)}, first={Roche-lobe overflow (RLOF)}}

\newglossaryentry{mchan}{name={Chandrasekhar mass~}, text={Chandrasekhar~mass}, symbol={\ensuremath{M_\textrm{Chan}}}, plural={Chandrasekhar~masses}, description={Mass when the core of a star collapses due to insufficient degeneracy pressure - for a white dwarf $\approx1.38\,M_\odot$ see \citet{1931ApJ....74...81C}}, first={Chandrasekhar~mass \citep[$M_\textrm{Chan}=1.38\,M_\odot$;][]{1931ApJ....74...81C}}, sort=mchan}

\newglossaryentry{w7}{name={W7 model},description={W7 model \citep{1984ApJ...286..644N}},first = {W7 model \citep{1984ApJ...286..644N}}}

\newglossaryentry{stats.pdf}{name=PDF, description={Probability Density Function}, first={Probability Density Function}}

\newglossaryentry{dpr}{name=DPR, description={Distributed Peer Review}, first={Distributed Peer Review (DPR)}}

\usepackage[textsize=footnotesize,color=white,bordercolor=red,linecolor=red]{todonotes}
\let\newtodo\todo\def\todo#1{\newtodo[inline]{#1}}

\begin{document}

\submitjournal{ApJ}
\shorttitle{\textsc{Probabilistic Estimation of SN~2002bo Explosion Parameters}}
\shortauthors{O'Brien et al.}

\title{Probabilistic Reconstruction of Type Ia Supernova SN~2002bo}
\correspondingauthor{John T. O'Brien}
\email{jobrien585@gmail.com, obrie278@msu.edu}
\author[0000-0003-3615-9593]{John T. O'Brien}
\affiliation{Department of Physics and Astronomy, Michigan State University, East Lansing, MI 48824, USA}

\author[0000-0002-0479-7235]{Wolfgang E. Kerzendorf}
\affiliation{Department of Physics and Astronomy, Michigan State University, East Lansing, MI 48824, USA}
\affiliation{Department of Computational Mathematics, Science, and Engineering, Michigan State University, East Lansing, MI 48824, USA}

\author[0000-0001-7343-1678]{Andrew Fullard}
\affiliation{Department of Physics and Astronomy, Michigan State University, East Lansing, MI 48824, USA}

\author[0000-0003-2544-4516]{Marc Williamson}
\affiliation{Department of Physics, New York University, New York, NY, 10003, USA}

\author{R\"udiger~Pakmor}
\affiliation{Max-Planck-Institut f\"ur Astrophysik, Karl-Schwarzschild-Str. 1, 85748 Garching, Germany}

\author[0000-0003-0426-6634]{Johannes Buchner}
\affiliation{Max-Planck-Institut f\"{u}r extraterrestrische Physik, Giessenbachstrasse 1, 85748 Garching bei M\"{u}nchen, Germany}

\author{Stephan Hachinger}
\affiliation{Leibniz Supercomputing Centre, Boltzmannstr. 1, 85748 Garching bei M{\"u}nchen, Germany}

\author[0000-0002-7941-5692]{Christian Vogl}
\affiliation{Max-Planck-Institut f\"ur Astrophysik, Karl-Schwarzschild-Str. 1, 85748 Garching, Germany}
\affiliation{Exzellenzcluster ORIGINS, Boltzmannstr. 2, 85748 Garching, Germany}

\author[0000-0002-8094-6108]{James H. Gillanders}
\affiliation{Astrophysics Research Centre, School of Mathematics and Physics, Queen's University Belfast, BT7 1NN, UK}

\author{Andreas Fl{\"o}rs}
\affiliation{GSI Helmholtzzentrum f\"ur Schwerionenforschung, Planckstra\ss e 1, 64291 Darmstadt, Germany}

\author[0000-0003-4418-4916]{Patrick van der Smagt}
\affiliation{Machine Learning Research Lab, Volkswagen AG, Munich, Germany}
\affiliation{Faculty of Informatics, E\"otv\"os Lor\'and University, Budapest, Hungary}



\begin{abstract}
    
    Manual fits to spectral times series of Type Ia supernovae have provided a method of reconstructing the explosion from a parametric model but due to lack of information about model uncertainties or parameter degeneracies direct comparison between theory and observation is difficult.
	In order to mitigate this important problem we present a new way to probabilistically reconstruct the outer ejecta of the normal Type Ia supernova SN~2002bo.  
    A single epoch spectrum, taken \num{10}\,days before maximum light, is fit by a \num{13}-parameter model describing the elemental composition of the ejecta and the explosion physics (density, temperature, velocity, and explosion epoch).  
    Model evaluation is performed through the application of a novel rapid spectral synthesis technique in which the radiative transfer code, \glstext{tardis}, is accelerated by a machine-learning framework.
    Analysis of the posterior distribution reveals a complex and degenerate parameter space and allows direct comparison to various hydrodynamic models. 
    Our analysis favors detonation over deflagration scenarios and we find that our technique offers a novel way to compare simulation to observation.  

\end{abstract}


\section{Introduction}
\label{sec:introduction}

\glspl{snia} are a spectral class of supernovae defined by their lack of hydrogen lines and the presence of silicon lines. \glspl{snia} are caused by the thermonuclear explosion of carbon-oxygen white dwarfs in binary systems forming a large amount of $^{56}$Ni, which drives the behavior of their light curves \citep{Colgate1969}.  They contribute significantly to the chemical evolution of their host galaxies through the dispersion of iron-peak elements formed during the explosion \citep[][see Figure 39]{Kobayashi2020}. 
 
Their ability to act as standardizable candles \citep{Phillips1993} has served as a powerful tool in constraining cosmological parameters \citep{Branch1992Hubble, Riess1998}, though there remains significant variation in their brightness that is unaccounted for \citep[e.g.][]{Blondin2012}. Furthermore, the identification of the ignition mechanism leading to \glspl{snia} remains an area of active research \citep[see e.g.][]{Polin2019}. 

The community has identified multiple promising pathways to explosions, many of which originate in a binary system.  For example, nuclear burning may be ignited by either the merger of two CO white dwarfs \citep[e.g.][]{Nomoto1982b, Webbink1984, Iben1984, van_Kerkwijk_2010, Livio_2003, Kashi2011}, or accretion from a companion star forming a near-Chandrasekhar mass CO white dwarf causing a central ignition \citep[e.g.][]{Whelan1973}, or accretion of a helium layer onto a sub-Chandrasekhar mass white dwarf \citep[e.g.][]{Woosley1994, Fink2010, Shen2018, Polin2019} leading to a surface helium detonation that propagates inward triggering central ignition.

Various models have been proposed to describe the processes underlying \glspl{snia}.  In particular, the speed at which the nuclear burning propagates through the star remains poorly understood. 
Reconstructing the explosion from spectral time series (also known as abundance tomography) is a crucial tool to understand the explosion scenario \citep[see e.g.][]{Mazzali2007}. Previous work into abundance tomography \citep[e.g.][]{Stehle2005,SAUER2008} has begun to show us a picture of how \gls{snia} explosions compare to theoretical models, but they lack a probabilistic interpretation of their parameters.

\noindent 

SN~2002bo is a ``Branch normal'' \citep{Branch1993, Benetti2004, Branch_2006} \gls{snia} discovered in NGC 3190 that has been modeled extensively in the literature \citep[e.g.][]{Stehle2005, SAUER2008, Benetti2004, Kerzendorf2011}.  Specifically, \citet{Stehle2005} used a multi-line Monte-Carlo code to manually reconstruct the explosion mechanism using \num{13} epochs of spectra.  Their inference suggests a Type Ia with moderate amounts of mixing of $^{56}$Ni and intermediate-mass elements, as well as a lack of carbon in the ejecta, indicating a possible explosion asymmetry and orientation effects.  

While these results offer a good foray into the investigation of the abundance tomography of \glspl{snia}, the lack of uncertainty or error analysis limits our ability to constrain the range of possible explosion scenarios.  Physical sources of uncertainty such a line-blending as well as potential parameter degeneracies
warrant the need for probability distributions.  

In this work, we present a method of Bayesian inference of supernova parameters by applying the radiative transfer code \gls{tardis}, accelerated by a machine-learning framework \citep{Dalek2020}, to a single spectrum of SN~2002bo taken \num{10}\,days before maximum light \citep{Benetti2004}. We begin with a description of our model and associated parameters in Section~\ref{sec:explosion_model}.  The sampling of the parameter space, including a discussion on prior distributions and resulting posterior distributions, is given in Section~\ref{sec:parameter_inference}. A summary of results can be found in Section~\ref{sec:results}.  Appendices are included to provide general background on the techniques used for spectral synthesis acceleration as well as additional data used in our analysis.  In Appendix~\ref{sec:emulator}, we outline a machine-learning framework used to accelerate \gls{tardis} evaluation.  Finally, in Appendix~\ref{sec:additional_figures}, links to data sources and data products are provided in order to assist researchers who wish to replicate our findings.

\section{Explosion Model}
\label{sec:explosion_model}

The optical spectrum of SN~2002bo \num{10}\,days before maximum light is modeled with spectral synthesis produced by the radiative transfer code \gls{tardis}.
\gls{tardis} is a modular framework that allows for the use of various physics modules and has been widely used for modeling a range of photospheric SNe \citep[e.g.][]{Magee2016, Boyle2017, Barna2017, Vogl2020, Gillanders2020, Williamson2021}. \gls{tardis} approximates the radiation field in the ejecta with an optically thick inner boundary and an optically thin homologously expanding ejecta above. There is no energy generation in the simulation area and the energy injection is purely set by the temperature, $T_\textrm{inner}$, and radius, $r_\textrm{inner}$, of this inner boundary. The optically thin ejecta is divided into a series of concentric shells in velocity space.  The velocity of each shell is determined by the inner boundary velocity, $v_\textrm{inner}$, and increases linearly up to an outer velocity boundary.  The radius of the inner boundary, $r_\textrm{inner}$, and consequently the radius of the shells, are set by the product of $v_\textrm{inner}$ with the time since the explosion, $t_\textrm{exp}$.

We employ a power law relationship of the density with the velocity parameterized by the power law index $\alpha_{\rho}$ such that $\rho_{\mathrm{shell}} \propto v_{\mathrm{shell}}^{\alpha_{\rho}}$\footnote{The reference density is pre-computed from the power law index to match that of the W7 model at \num{10000}\,km/s}.  In previous works \citep{Stehle2005, Kerzendorf2011}, the density profile of \gls{snia} ejecta has been described by a 1-dimensional parameterized explosion model known as W7 \citep[see e.g.][]{Nomoto1984III} which can be approximated as a power law between velocity and density with an exponent of \num{-7} \citep{Branch1985}.  In order to account for deviations from the W7 power law profile we have left the power law index as a free parameter in our study, the prior for which can be found in Table~\ref{tab:prior_posterior}.

We approximate the elemental composition of the ejecta by assuming a uniform distribution of abundances above the photosphere (the same abundance values are used in each shell).  
We explored a set of abundances commonly used in the literature \citep[e.g.][]{Stehle2005,SAUER2008,Kerzendorf2011}, namely carbon, magnesium, silicon, sulfur, calcium, titanium, and chromium.  Iron, cobalt, and nickel abundances were split up into the decay chain of the isotope $^{56}$Ni and stable iron. 
These elements account for the majority of the mass in explosion models and are well constrained by the spectra of \glspl{snia} \citep{Filippenko1997}.  
The set of abundances (C, Mg, Si, S, Ca, Ti, Cr, Fe$_{\textrm{stable}}$, and $^{56}$Ni) and explosion parameters ($T_{\textrm{inner}}$, $v_\textrm{inner}$, $t_\textrm{exp}$, and $\alpha_{\rho}$) all together compose a 13-dimensional parameter space to model our spectra.

For the plasma state, we have chosen the \texttt{nebular} ionization approximation implemented in \gls{tardis} and the \texttt{dilute-lte} excitation approximation. The radiation-matter interaction is modeled using the \texttt{macroatom} prescription. We have also set the number of packets to be equal to \num{400000}. The final spectral calculation uses the formal integral method \citep{Lucy1999B} rather than straight packet statistics.  Configuration of \gls{tardis} can be found in Appendix~\ref{sec:additional_figures}.

\subsection{Model Evaluation}
\label{sec:model_evaluation}

Spectral synthesis from our model with \gls{tardis}, on average, takes approximately 10 minutes of CPU time on an Intel\textsuperscript{\textregistered} Xeon\textsuperscript{\textregistered} E5-2670 v2 CPU.  \citet{Dalek2020} estimates the time required to explore a 20-parameter toy-model at this rate to be $\sim$\num{420}\,years.  Such a time constraint on model evaluation imposes a restriction upon our ability to use radiative transfer codes as a method of exploring the posterior distribution of \gls{snia} models.  In order to subvert this restriction, we have implemented a technique for speeding up our model evaluation by \num{8} orders of magnitude based upon the machine-learning framework developed by \citet{Dalek2020}.  The estimation of our models through this technique is known as {\em emulation} and the machine-learning framework we used will from here on be referred to as the {\em emulator}.  Details of the emulator including architecture, accuracy, and error analysis can be found in Appendix~\ref{sec:emulator}.  We find our emulator predicts the synthetic spectra produced by \gls{tardis} given a set of model parameters within \num{1}\% and is therefore an effective and necessary substitute for model evaluation.


\section{Parameter Inference}
\label{sec:parameter_inference}
Vectors of candidate input abundances (carbon, magnesium, etc.) and explosion parameters, $\vec{\theta}=\{\textrm{C}, \textrm{Mg}, \ldots, t_\textrm{exp}, \alpha_{\rho}\}$, are drawn from a prior-distribution described in Section~\ref{sec:prior_distribution}.  
Model spectra are then produced by the emulator, where the emulated synthetic spectrum is predicted using the input parameters $\vec{\theta}$.  We determine the likelihood of a given model through the application of a likelihood function described in Section~\ref{sec:likelihood_estimation}.  We have developed a non-$\chi^{2}$ likelihood function that takes into account systematic differences between our theoretical and observed spectra.  Lastly, in Section~\ref{sec:posterior_distribution}, we outline the Monte Carlo sampling technique used to construct the posterior distribution.

\subsection{Prior Distribution}
\label{sec:prior_distribution}

We developed a distribution from which to draw our prior samples based on parameters of \gls{snia} abundances taken from the Heidelberg Supernova Model Archive (\textsc{HESMA}).  We specifically used the set of abundance profiles provided from various \glspl{snia} hydrodynamic simulations \citep{2014MNRAS.438.1762F,2017MNRAS.472.2787N,2013MNRAS.429.2287K,2015MNRAS.450.3045K,2010ApJ...714L..52S,2017MNRAS.472.2787N,2018A&A...618A.124F,2015A&A...580A.118M,2010A&A...514A..53F,2010ApJ...719.1067K,2012MNRAS.420.3003S,2020A&A...635A.169G} to determine the range of input parameters.  We determined the bounds of our prior by taking the 60\% quantile of the distribution of abundances from the \textsc{HESMA} models where the shell velocity was above \num{10000}\,km/s in order to be consistent with the expected structure of the outer shells.  

Abundances were sampled uniformly 
in log-space with any remaining abundance fraction filled in with oxygen such that all abundance fractions summed to unity.  Oxygen is often used as a ``filler'' element in supernova fitting \citep[e.g.][]{2017MNRAS.471..491H} due to the insensitivity to changes in the spectrum with respect to the oxygen mass fraction \citep[cf.][Sec. 2.2.5.2]{hachinger_phd2011}.  The oxygen abundance is therefore only determined implicitly and is not included as a model parameter.

For all other model parameters, we sampled along a uniform distribution.  We used the values for explosion time, ejecta velocity, photospheric boundary temperature, and density profile power law exponent from the fit made by \citet{Kerzendorf2011} as centroids.  We then reviewed the works of \citet{Stehle2005} and \citet{Benetti2004} to determine reasonable ranges of uncertainties on these values which were used to set the edges of the distribution.  The range of values sampled for each parameter can be found in Table~\ref{tab:prior_posterior}.




\subsection{Likelihood Estimation}
\label{sec:likelihood_estimation}

While our emulator accurately recreates the behavior of \gls{tardis} under our spectral synthesis model, observations of real spectra are subject to physical and systematic biases.  In order to compare our model spectra, $\hat{f}(\vec{\theta})$, to observation, $f_{\rm obs}$, we develop a likelihood function, $\mathcal{L}(\vec{\theta})$, that corrects our model spectra and compares the results to our observed spectrum.  

A correction function, $C(\hat{f}(\vec{\theta}))$, is applied to our model spectra.
$C(\hat{f}(\vec{\theta}))$ first applies a redshift correction to set the frame of the model spectrum to the observed frame of SN~2002bo at z=\num{0.0042} \citep{Benetti2004}.  
A host extinction correction is then performed using the model described by \citet{ccm89} using $R_{V}=3.1$ \citep{Schlafly2011} and $E(B-V)=0.3$ \citep{Benetti2004}.  Finally, a continuum removal technique described by \citet{Tonry1979} and \citet{Blondin2007} is applied to the model spectrum.    The continuum is estimated using a zero-mean 13-point cubic spline fit to the spectrum.  We apply this continuum removal to our model spectra first, then we multiply by the continuum that would be removed by applying the same technique to the observed spectrum.  Finally, the resulting continuum-removed model spectrum is linearly interpolated to the wavelength bins of the observed spectrum.  Applying the corrections in this way allows us to compare our simulated spectra directly to the observed spectrum. 

We compare our corrected model spectrum to the observed spectrum using a Gaussian likelihood function,


$$\log\mathcal{L}(\vec{\theta}) = -\frac{1}{2}\sum_{\lambda}\left[\frac{(C(\hat{f}(\vec{\theta})) - f_\mathrm{obs})_{\lambda}^{2}}{s^{2}} + \log (2\pi s^{2}) \right],$$



\noindent where $\lambda$ represents the wavelength bin of the observed spectrum of SN~2002bo in the observed frame.  The parameter $s^{2}$ estimates the variance of our posterior distribution over model spectra which we infer as another parameter \citep{Hogg2010} with a log-uniform prior.  



\subsection{Posterior Distribution}
\label{sec:posterior_distribution}

The topology of the posterior distribution is unknown a priori, and could contain complicated degeneracies or multimodalities. Nested sampling \citep{Skilling2004, Buchner2021} is a robust Monte Carlo technique for this setting. We use the MLFriends algorithm \citep{Buchner2014, Buchner2017} implemented in the {\em UltraNest} package \citep{BuchnerJoss2021}.
The posterior distribution was explored with \num{400} live points. It converged to the target distribution after \num{10000} iterations and required \num{1000000} model evaluations.




\section{Results}
\label{sec:results}




\begin{figure*}
    \centering
    \includegraphics[width=0.9\linewidth]{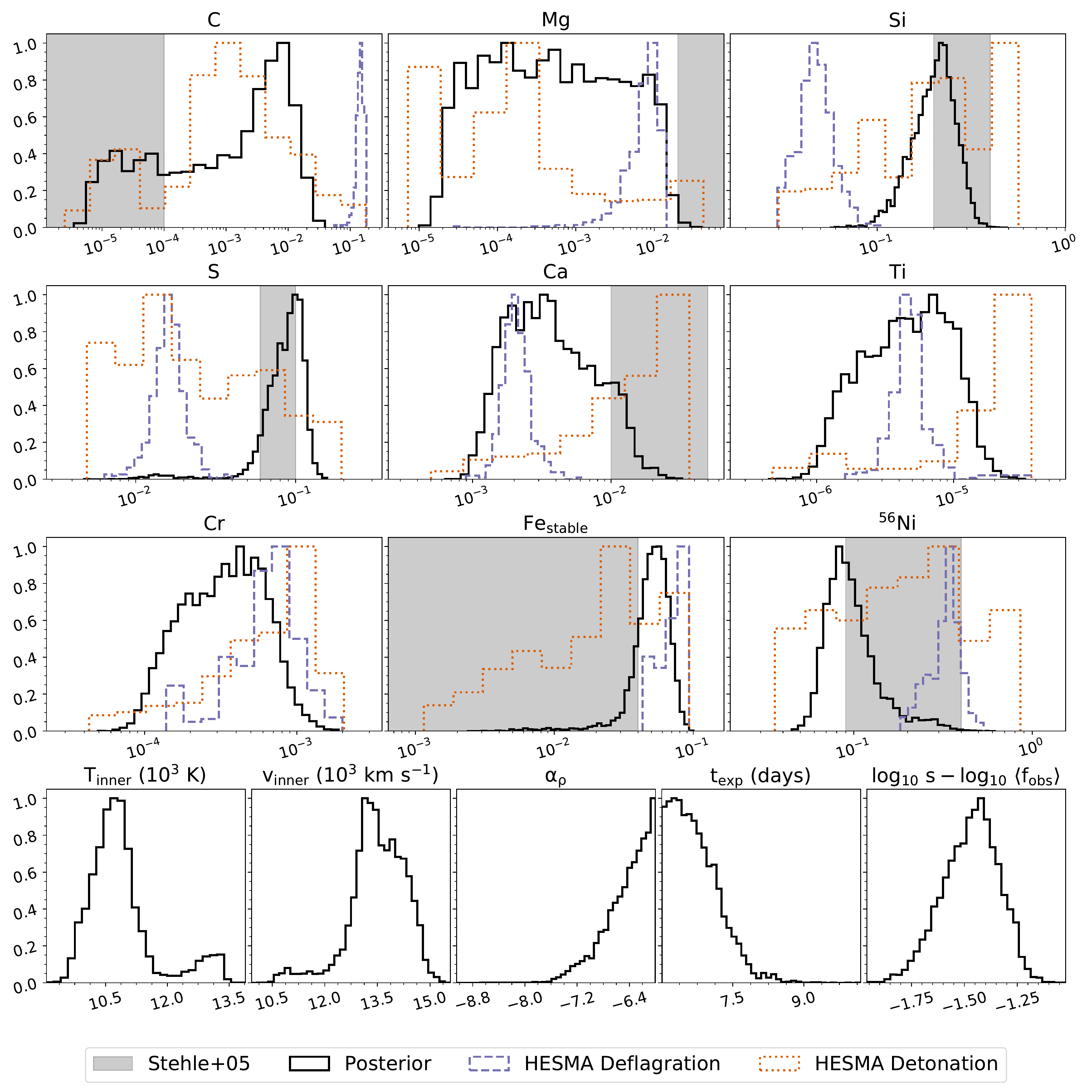}
    \caption{\label{fig:posterior} Posterior distribution of the parameter space sampled using nested sampling (Black).  Overlaid are distributions of elemental abundances above \num{10000}\,km/s taken from various \textsc{HESMA} models.  Pure deflagration models are shown in green while pure detonation models are shown in orange.  DDT models are not included as they would not be noticeably distinguishable from pure detonation models at this early epoch.  Estimates of the range of abundances of elements in ejecta layers between \num{10000}\,km/s and \num{15000}\, km/s from \citet{Stehle2005} are represented by the grey shaded regions.  Due to differences in methodology we do not have reliable estimates for the abundaces of titanium and chromium from \citet{Stehle2005}.
    }
\end{figure*}

Figure~\ref{fig:posterior} shows the converged parameter distributions from our statistical inference.  Silicon and sulfur abundances contribute the largest fraction by mass of the ejecta which can be inferred from the spectral features present in SN~2002bo.  
\citet{Stehle2005} used a similar code to \gls{tardis} to manually fit the spectral time series of SN~2002bo. However, due to differences in methodologies, direct comparison of elemental abundances is difficult and must be approximated.  Since \citet{Stehle2005} does not provide uncertainties, we make the assumption that the uncertainty in their reported elemental abundances within various layers of the ejecta are comparable to those found in our study.  
Unfortunately, the full model inferred by \citet{Stehle2005} is not directly available for download so we estimate abundances in terms of mass fractions from the figures \citep[][Figure 5]{Stehle2005}.  

\begin{table*}
    \centering
    \begin{tabular}{lrrrrr}
    \multicolumn{1}{c}{Parameter} & \multicolumn{2}{c}{Prior Bounds} & \multicolumn{3}{c}{Posterior Percentiles}\\
    & Minimum & Maximum & 16\% & 50\% & 84\% \\
    \hline
    $\mathrm{C}$ & $2.3\times 10^{-6}$ & $0.17$ & $9.5\times 10^{-5}$ & $0.0015$ & $0.0085$ \\
    $\mathrm{Mg}$ & $8.3\times 10^{-6}$ & $0.036$ & $0.00011$ & $0.00049$ & $0.0047$ \\
    $\mathrm{Si}$ & $0.029$ & $0.58$ & $0.17$ & $0.21$ & $0.26$ \\
    $\mathrm{S}$ & $0.005$ & $0.19$ & $0.074$ & $0.09$ & $0.11$ \\
    $\mathrm{Ca}$ & $0.00043$ & $0.039$ & $0.0021$ & $0.0034$ & $0.0084$ \\
    $\mathrm{Ti}$ & $4.4\times 10^{-7}$ & $3.7\times 10^{-5}$ & $2.7\times 10^{-6}$ & $4.7\times 10^{-6}$ & $9.7\times 10^{-6}$ \\
    $\mathrm{Cr}$ & $3.8\times 10^{-5}$ & $0.0022$ & $0.00021$ & $0.00034$ & $0.00062$ \\
    $\mathrm{Fe_{stable}}$ & $0.0011$ & $0.1$ & $0.044$ & $0.052$ & $0.065$ \\
    $\mathrm{^{56}Ni}$ & $0.037$ & $0.85$ & $0.078$ & $0.091$ & $0.13$ \\
    \hline
    $T_{\textrm{inner}}\ (\mathrm{K})$ & \num{8000} & \num{18000} & $10383$ & $10720$ & $11357$ \\
    $v_{\textrm{inner}}\ (\mathrm{km}\ \mathrm{s}^{-1})$ & \num{7000} & \num{20000} & $13100$ & $13508$ & $14291$ \\
    $\alpha_{\rho}$ & \num{-10} & \num{-6} & $-6.10$ & $-6.36$ & $-6.63$ \\
    $t_{\textrm{exp}}\ (\mathrm{days})$ & \num{6} & \num{13} & $6.32$ & $6.64$ & $7.21$ \\
    \hline
    $\mathrm{\log_{10} s}$ & $-18$ & $-14$ & $-15.91$ & $-15.81$ & $-15.69$
\end{tabular}
    \caption{\label{tab:prior_posterior}The range of parameters sampled from our prior distribution along with their estimates determined by the posterior distribution.  The abundance distributions are based upon log-uniform sampling but modifications are made in order to assure that the sum of abundance parameters add to unity.  All other values displayed are sampled uniformly.  For a full description of the abundance sampling method see Section~\ref{sec:prior_distribution}. Elemental abundances are shown in terms of mass fractions.  Estimates from the posterior distribution are presented as the median with the edges of the \num{68}\% confidence interval.} 
\end{table*}

We compare our findings to their range of abundances reported in the velocity interval from \num{10000}\,km/s to \num{15000}\,km/s and generally find good agreement within our uncertainty ranges. 
We find a significant lack of carbon in the ejecta consistent with their analysis.  The range of abundances determined from their analysis of silicon (0.2 - 0.4), sulfur (0.06-0.1), and $^{56}$Ni (0.09 - 0.11) all overlap with our 68\% confidence interval in Table~\ref{tab:prior_posterior}.  Their abundances of iron ($<10^{-4}$ - 0.04) and calcium (0.01 - 0.05) were slightly outside this region but are consistent if the level of uncertainty in their analysis is similar to ours. Individual values for both titanium and chromium are not available so performing a direct comparison is not particularly reasonable or reliable.  

By far our largest deviation from \citet{Stehle2005} is our magnesium abundance.  Magnesium has the largest range of uncertainty in our analysis, spanning nearly four orders of magnitude.  Operating under the assumption that the uncertainties in \citet{Stehle2005} are comparable to ours, not much information can be gathered from a comparison of values between the two studies as the magnesium abundance is mostly uninformative. 

We constrain $t_{\textrm{exp}}=6.64^{7.21}_{6.32}$\,days\footnote{See Table~\ref{tab:prior_posterior} for description of quantification} which is slightly below that of \citet[][$t_{\textrm{exp}}=$\,\num{7.9}\,$\pm$\,\num{0.5}\,days]{Benetti2004} and \citet[][$t_{\textrm{exp}}=$\,\num{8.04}\,days]{Stehle2005}.  Our estimates for both $T_{\textrm{inner}}$ and $v_{\textrm{inner}}$ are consistent with the range of values found by \citet{Stehle2005} for spectra between nearby epochs.  The overall agreement of our results with similar previous attempts at manual fitting as well as theoretical models for \glspl{snia} explosion physics demonstrates that our model is consistent with the current literature.

\begin{figure}
    \includegraphics[width=1.0\linewidth]{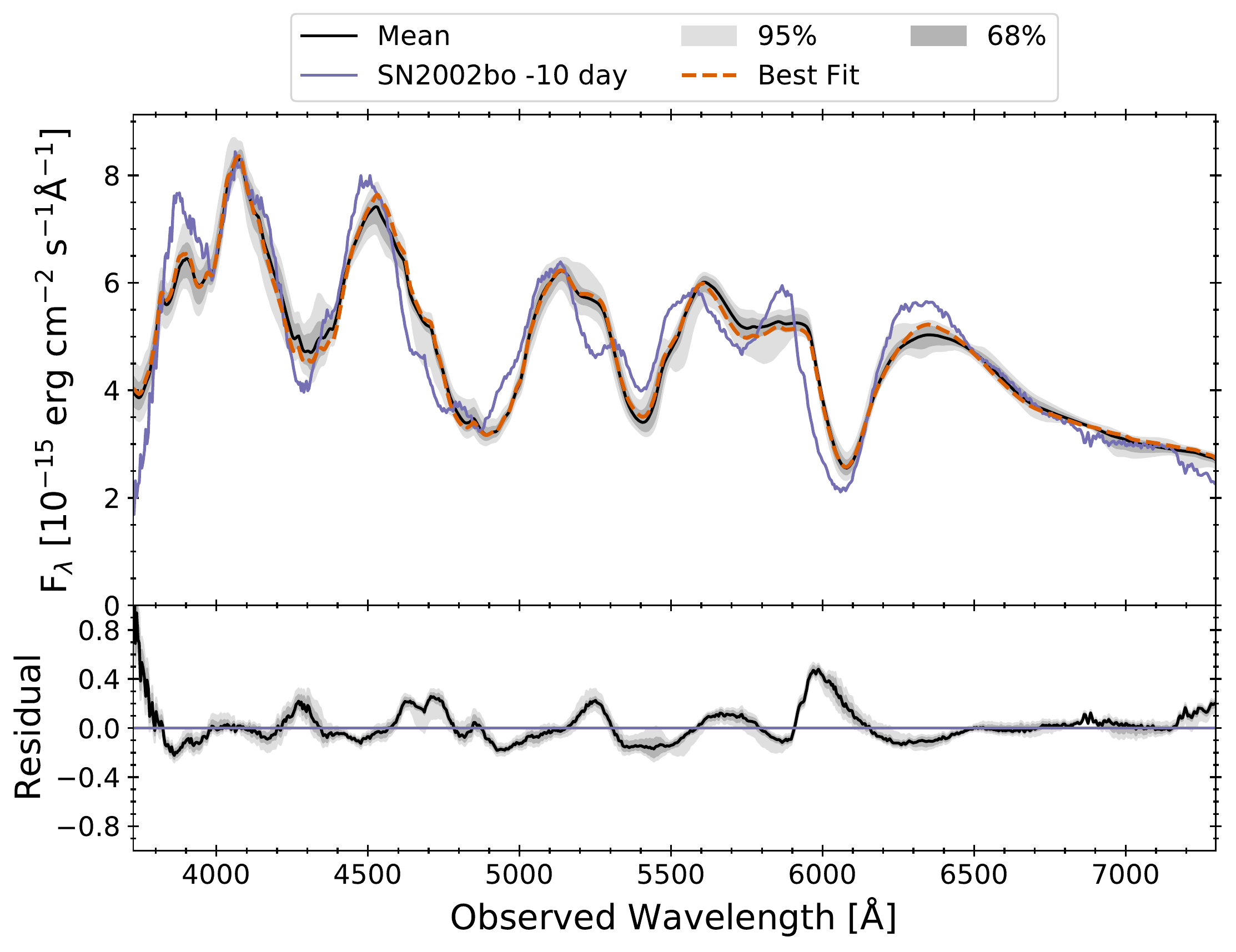}
	\caption{\label{fig:spectrum} Fit to observed SN~2002bo \num{-10} day spectrum (green) using nested sampling to sample the posterior distribution.  The best fit spectrum (orange), represented by the maximum likelihood sample, shows a decent fit to the spectrum but misses features around \num{5972}\,\AA\ and \num{3900}\,\AA\ as well as much of the UV.  The mean of the posterior distribution is shown in black with the \num{68}\% and \num{95}\% regions  in grey and light grey respectively.  Posterior spectra are presented after application of the correction function described in Section~\ref{sec:likelihood_estimation}.  The residual distribution is shown as the fractional error between our posterior and our observed spectrum.
	}
\end{figure}

There are a few notable mismatches between our posterior spectra and the observed spectrum (Figure~\ref{fig:spectrum}).  In the \ion{S}{2} doublet our model over-fits the left peak and under-fits the right peak.  This discrepancy is a common occurrence in radiative transfer model fits \citep[see e.g.][]{Stehle2005} to \gls{snia} spectra and is due to a poor understanding of the lines lists and occupation numbers in this region.  Since our abundance distribution through the ejecta is approximated to be uniform, the iron abundance in the outer layers is generally overestimated.  This causes line blanketing as the bluer packets are reflected back inwards resulting in a higher radiative temperature as well as less flux at the blue end of the spectrum.  The higher temperatures affect the overall ionization state of the plasma causing the \ion{Si}{2} to \ion{Si}{3} ratio to decrease, weakening the \ion{Si}{2} (\num{5972}\, \AA) feature.  The poor fit to the \ion{Si}{2} doublet is also seen in previous studies \citep[see e.g.][]{Benetti2004}.

We are able to perform a direct comparison of inferred model parameters of a real \gls{snia} spectrum to statistical samples of theoretical explosion models.  
In addition to the posterior distributions of the model parameters inferred for SN~2002bo, Figure~\ref{fig:posterior} shows the distribution of abundances from two classes of models taken from the \textsc{HESMA} data sets above \num{10000}\,km/s corresponding to pure-deflagrations and pure-detonations.  Deflagration to detonation transition (DDT) models are not included as they would be indistinguishable from pure-detonation models above the photosphere at these early times.  
The posterior distribution best matches with the distribution of abundances sampled from the \textsc{HESMA} detonation models, while mostly excluding the pure deflagration models.  The unfavorability of pure-deflagration models is strongly apparent for the distribution of carbon, sulfur, and silicon abundances in Figure~\ref{fig:posterior}.  Calcium and chromium abundances slightly favor pure-deflagration hydrodynamic models, though their distribution widths are large and stretch over a few orders of magnitude indicating that these abundances are not affecting the final shape of the spectrum significantly.  
We find that our initial modeling of the \num{-10}\,day spectrum of SN~2002bo generally favors detonation or DDT models. 

\begin{figure}
        \centering
        \includegraphics[width=1.0\linewidth]{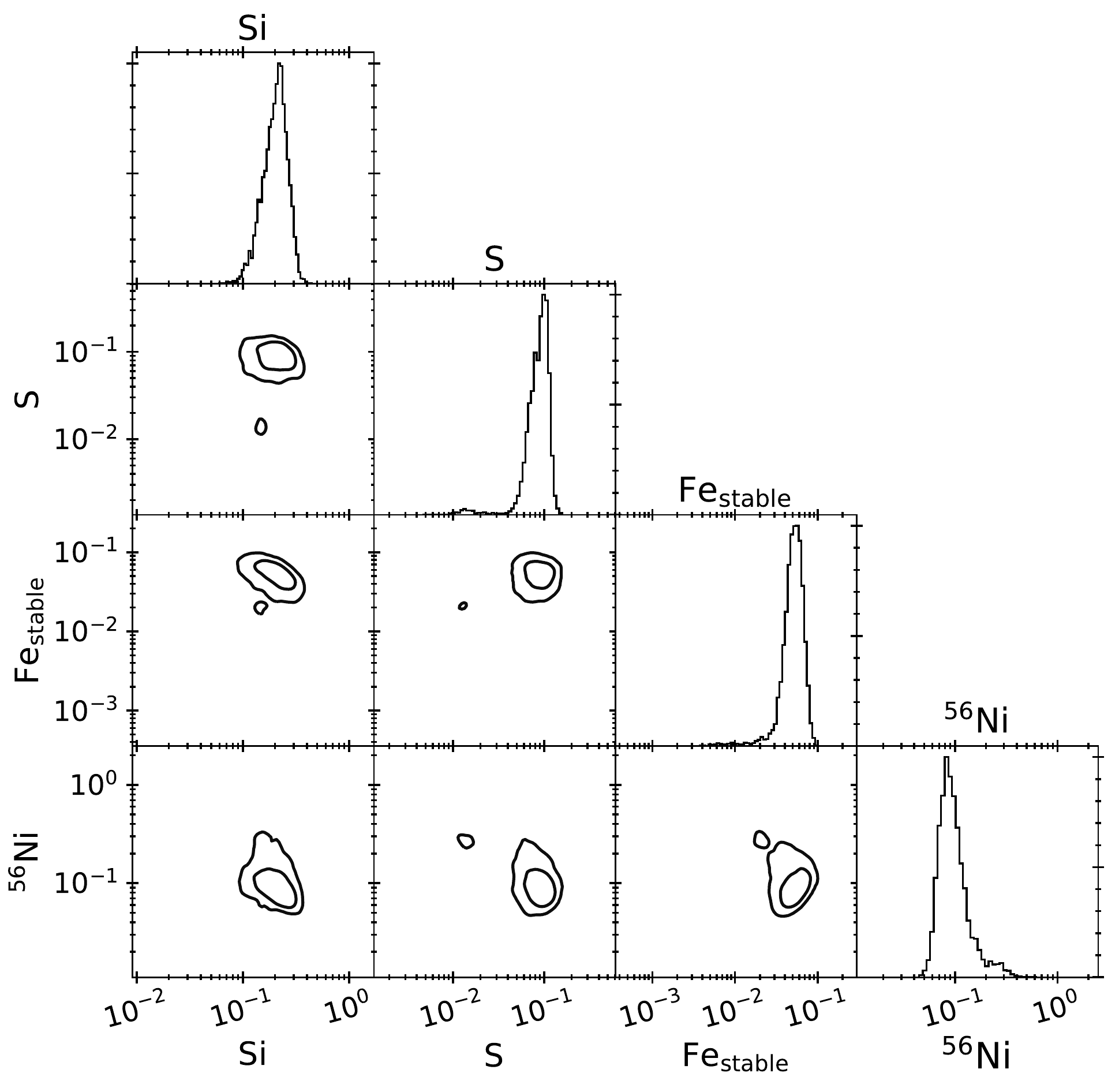}
        \caption{\label{fig:degen} Posterior probability distribution of the elemental abundances of silicon, sulfur, stable iron, and $^{56}$Ni. Contours show \num{68}\% and \num{95}\% confidence intervals of the Gaussian kernel density estimation (KDE) over the joint distribution of each parameter.  Degeneracies and multimodalities in elemental abundances are apparent.}
\end{figure}

Figure~\ref{fig:degen} demonstrates the complexity of the posterior distribution of elemental abundances.  
A small multimodality in the sulfur abundance raises the possibility of manual fits becoming trapped in local minima.
The joint probability distribution of stable iron with both silicon and $^{56}$Ni is degenerate and multimodal.
Such complexities indicate that any single set of model parameters may only describe one of a distribution of parameters that all appear to model the observed spectrum to similar accuracy.  
Despite some of the large variations and complexity in the posterior distribution of parameters (Figure~\ref{fig:posterior}), the distribution of model spectra produced by these parameters (Figure~\ref{fig:spectrum}) is within \num{3}\% variation of the mean of the observed spectrum.  

\section{Conclusion}
\label{sec:conclusion}

We present a probabilistic reconstruction of a \gls{snia} explosion. Our results generally agree with manual fits \citep[see e.g.][]{Stehle2005}.  
We estimate the distribution of elemental abundances required to reproduce the observation of an early-time spectrum of SN~2002bo.  
Degeneracies and multimodalities in certain parameters showcase the need for a Bayesian treatment to draw secure physical conclusions since similar spectra may be synthesized over a wide and complex space of parameters.   
The posterior distribution is compared to the distribution of elemental abundances computed from various explosion models in \textsc{HESMA}.  We find that our analysis favors detonation models over pure-deflagration models.  Given the speed and effectiveness of our modeling technique, we have demonstrated a new avenue for investigating the inner mechanisms driving \gls{snia} explosions.



\section{acknowledgements}
We would like to thank Stuart Sim and Maryam Modjaz for their edits and suggestions.\\

This work was supported in part through computational resources and services provided by the Institute for Cyber-Enabled Research at Michigan State University.\\

This work made use of the Heidelberg Supernova Model Archive (\textsc{HESMA}), \url{https://hesma.h-its.org}\\

This research made use of \textsc{tardis}, a community-developed software package for spectral
synthesis in supernovae \citep{TARDIS2014, kerzendorf_wolfgang_2021_4995779}. The
development of \textsc{tardis} received support from the Google Summer of Code initiative,
from ESA's Summer of Code in Space program, and from NumFOCUS's Small Development Grant. \textsc{tardis} makes extensive use of Astropy
and PyNE.\\

This work would not have been possible without the large open source software community providing powerful numerical, scientific, visualization, machine-learning, and astrophysical libraries: Astropy\footnote{\url{https://www.astropy.org}} \citep{astropy:2013,astropy:2018}, extinction\footnote{\url{https://extinction.readthedocs.io/en/latest}} \citep{exctinction}, Matplotlib\footnote{\url{https://matplotlib.org}} \citep{matplotlib}, Numba\footnote{\url{https://numba.pydata.org}} \citep{numba}, NumPy\footnote{\url{https://numpy.org}} \citep{numpy}, pandas\footnote{\url{https://pandas.pydata.org}} \citep{pandas}, scikit-learn\footnote{\url{https://scikit-learn.org}} \citep{sklearn}, SciPy\footnote{\url{https://www.scipy.org/}} \citep{scipy}, 
TensorFlow\footnote{\url{https://www.tensorflow.org/}} \citep{tensorflow}, and UltraNest\footnote{\url{https://johannesbuchner.github.io/UltraNest}} \citep{Buchner2014,Buchner2019}.  We would like to thank these communities for providing and maintaining the resources that allow science to be done in an open, replicable, and accessible way.\\ 

C.V. was supported for this work by the Excellence Cluster ORIGINS, which is funded by the Deutsche Forschungsgemeinschaft (DFG, German Research Foundation) under Germany's Excellence Strategy -- EXC-2094 -- 390783311.

\bibliographystyle{aasjournal}
\bibliography{sources}

\section{Contributor Roles}
\begin{itemize}
\item Conceptualization:
John O'Brien, Wolfgang Kerzendorf
\item Data curation: John O'Brien
\item Formal Analysis:
John O'Brien, Wolfgang Kerzendorf
\item Funding acquisition:
Wolfgang Kerzendorf
\item Investigation: John O'Brien
\item Methodology:
John O'Brien, Wolfgang Kerzendorf
\item Project administration:
Wolfgang Kerzendorf
\item Resources: Institute for Cyber-Enabled Research at Michigan State University
\item Software:
John O'Brien,
Wolfgang Kerzendorf,
Marc Williamson,
Johannes Buchner, Christian Vogl,
James Gillanders,
\gls{tardis} Collaboration
\item Supervision:
Wolfgang Kerzendorf
\item Validation: John O'Brien
\item Visualization: 
John O'Brien,
James Gillanders
\item Writing – original draft:
John O'Brien,
Wolfgang Kerzendorf
\item Writing – review \& editing:
John O'Brien,
Andrew Fullard,
Marc Williamson,
Patrick van der Smagt,
Johannes Buchner,
James Gillanders,
Wolfgang Kerzendorf,
R\"udiger~Pakmor,
Andreas Fl{\"o}rs,
Stephan Hachinger,
Christian Vogl,
Stuart Sim,
Maryam Modjaz
\end{itemize}

\newpage
\begin{appendix}

\renewcommand\thefigure{\thesection.\arabic{figure}}
\section{Emulator}
\label{sec:emulator}
\setcounter{figure}{0}

Emulation is the practice of developing some analytic function that approximates the behavior of another function.  \gls{tardis} can be thought of as a function mapping a vector of supernova parameters to a vector representing a spectrum. We extend the techniques described in the \citet{Dalek2020} paper to make an emulator for the \num{-10}\,day spectrum of SN~2002bo.   The method proposed by \citet{Dalek2020} uses an ensemble of feed-forward neural networks to emulate the spectrum computation.  Our neural network is trained from a set of pre-computed data points, composed of training spectra over a grid spanning a physically plausible parameter space for a \gls{snia}.  The goal for the emulator is to be used in our parameter inference so we ensure that the training set parameter space contains the final prior fitting space (see Section~\ref{sec:prior_distribution}). 

We changed several parts of the procedure when compared to the emulator described by \citet{Dalek2020}.
One key difference is the addition of two parameters: the power law index $\alpha_\rho$ and the time since explosion $t_{\textrm{exp}}$.  
The bounds on parameters corresponding to computed spectra were also modified to encompass elemental abundances corresponding to shells above \num{8000}\,km/s in \textsc{HESMA} models.
\citet{Dalek2020} presented an ensemble of  different neural network architectures that could reproduce simulated \gls{tardis} spectra to a high degree of precision.  For this paper, for computational efficiency, we chose only a single network from the neural networks described by \citet{Dalek2020}.  Specifically, we used a model which propagates the \num{14} inputs through three subsequent hidden layers of \num{400} neurons each, reaching \num{500} outputs.  The hidden units used the ``softplus'' activation function.  We trained our emulator with the ``nadam'' optimizer on a \num{91000} sample training set and \num{39000} sample validation set in a \num{70}\%/\num{30}\% training/validation split.  Training time was \num{20} minutes on an NVIDIA\textsuperscript{\textregistered} GeForce\textsuperscript{\textregistered} RTX 2080Ti GPU.

\begin{figure}
    
    \begin{centering}
        \includegraphics[width=0.49\textwidth]{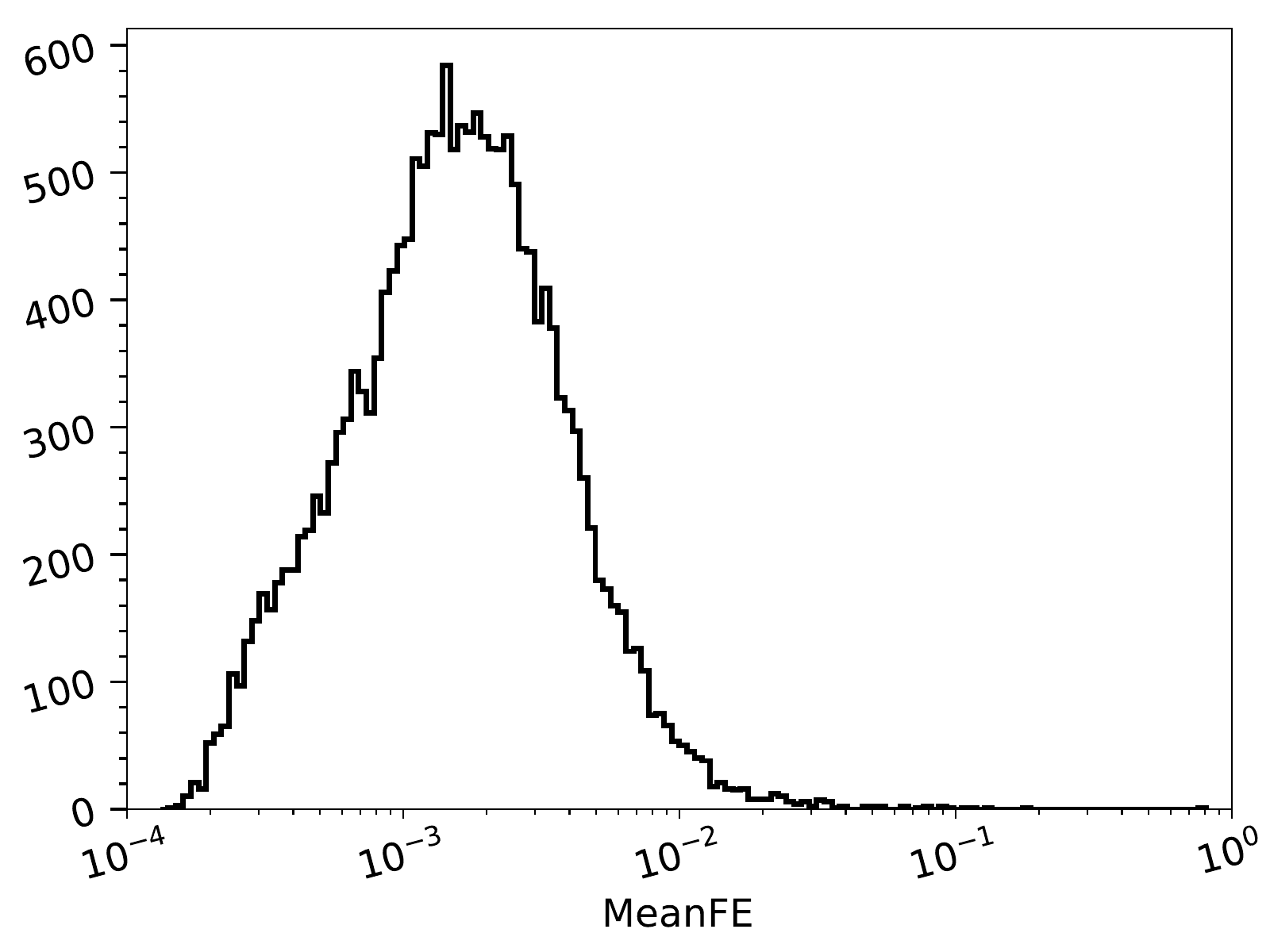}
        \includegraphics[width=0.49\textwidth]{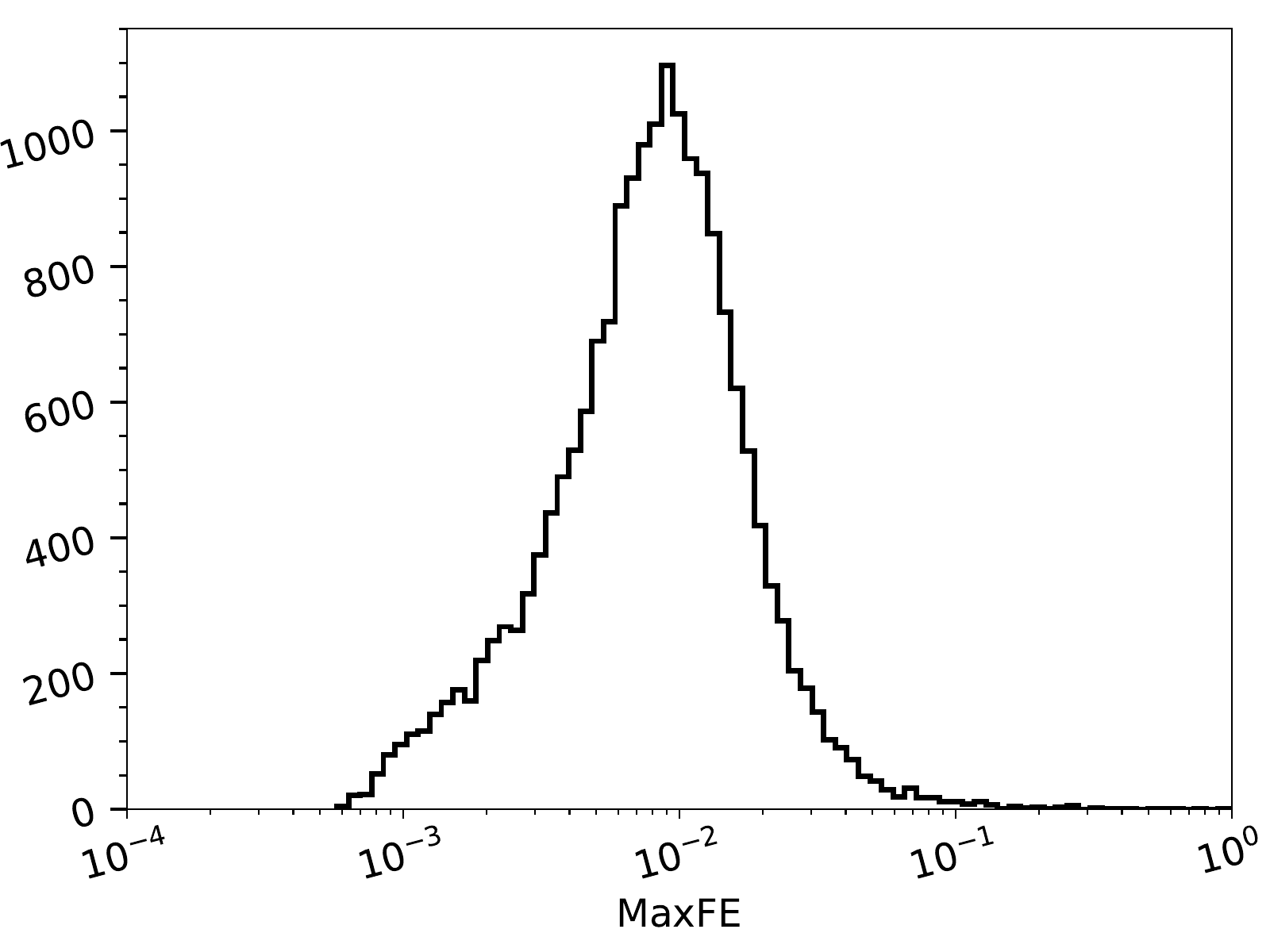}
        \caption{\label{fig:mean_max_fe} Mean and Maximum fractional error for our \gls{tardis} emulator.  Test spectra are compared to emulated spectra generated using the same parameter set.  The low level of error demonstrates that our emulator is effective at modeling the physics of \gls{tardis}.  Descriptions of the mean and maximum fractional error can be found in \citet{Dalek2020}}
    \end{centering}
    
\end{figure}

The measured accuracy of our emulator using the mean and maximum fractional error (Figure~\ref{fig:mean_max_fe}) is similar to that of the initial \textsc{dalek} emulator. Figure~\ref{fig:mean_max_fe} shows that our mean fractional error is almost always below 1\% over our validation set. The final fit presented in Section~\ref{sec:results} has a mean fractional error of \num{10}\% between the observed spectrum and the maximum posterior model indicating that any uncertainty from our emulation is less than systematics for the presented work.

\newpage
\section{External Links to Data}
\label{sec:additional_figures}


The \gls{tardis} configuration file, posterior samples with their associated weights, and the parameter grid and corresponding spectra used in training the emulator are provided through Zenodo: \dataset[10.5281/zenodo.5007378]{\doi{10.5281/zenodo.5007378}}.
The observed spectrum of SN~2002bo used in this paper is hosted by the \href{https://sne.space/}{Open Supernova Catalog} \citep{sne.space}.




\end{appendix}

\end{document}